\documentstyle[epsfig]{l-school}

\newcommand{\apj}{ApJ}
\newcommand{\mnras}{MNRAS}

\newcommand{\apjl}{ApJ (Letters)}

\begin{document}

\markboth{R. G. Abraham}{MORPHOLOGY} 
\setcounter{part}{1} 
\title{PERSPECTIVES IN PHYSICAL MORPHOLOGY} 
\author{R. G. Abraham} 
\institute{Institute of Astronomy, University of Cambridge, Madingley Road,
Cambridge CB3 OHA, United Kingdom} 
\maketitle

\section{INTRODUCTION}

In the first section of these lectures I outline the classical
framework of the Hubble classification system.  Because of space
limitations I will focus on points of controversy concerning the
physical interpretation of the Hubble sequence, showing how
morphological ideas shape our understanding of galaxy evolution.  I
will then present an overview of the remarkable progress made in
recent years in understanding how the local morphological composition
has transformed into that seen the distant Universe, highlighting work
from the {\em Hubble Deep Field} (HDF).  Recent studies show quite
clearly that the Hubble system does not provide a useful framework
for describing the appearance of galaxies at redshifts $z >1.5$. I
argue that as a result of this work the Hubble system needs to be
replaced by a system that is more objective, more physically
meaningful, and which is applicable across a wider range of redshifts.

\section{CLASSICAL MORPHOLOGY AND THE HUBBLE SYSTEM}

\subsection{How should we judge the Hubble system?}

As Sandage points out in the superb introduction to the {\em Carnegie
Atlas of Galaxies} \cite{Sandage:1994}, successful classification
systems can lead to enormous advances in science. Perhaps the best
example of this phenomenon is the periodic table, which played a
direct role in the development of models for atomic structure.

The central idea in any {\em physically motivated} (as opposed to
merely descriptive) classification system is that of {\em ranking}, in
the belief that this ranking reflects an underlying order in the basic
physics of the system, rather than merely reflecting the natural
tendency of the human brain to impose order and organization on
randomness in order to facilitate communication and memory. The choice
of features upon which the ranking is defined is, in the first
instance, completely subjective. In this stage of the process
intuition, experience, and inspiration are paramount (see the books by
Sandage \cite{Sandage:1994} and van den Bergh \cite{vdB:1998} for
insight into the creative processes through which galaxy
classification systems are developed).  But once the system is in
place, subjectivity should drop away as the system is tested against
the underlying physics. Therefore in my view an ideal classification
should have three characteristics:

\noindent (a) {\em physical significance:} The ranking imposed by the
classifier should track important underlying physical processes.

\noindent (b) {\em completeness:} There should be no gaping holes in
the system, ie. the vast majority of objects should slot naturally
into the system somewhere.

\noindent (c) {\em objectivity:} The system should be sufficiently
well-defined that classifications made independently by well-trained
observers should be very similar.

These are the metrics we will use when judging the success of the
Hubble system in the latter part of these lectures. My own verdict
will be given in \S2.6.

\subsection{Physical foundations of the Hubble sequence}

Figure 1 shows the familiar ``tuning fork'' diagram, representing
Sandage's definitive exposition \cite{Sandage:1961} of the system
described by Hubble in his book {\em The Realm of the Nebulae}
\cite{Hubble:1936}. I assume the reader is familiar with the basic
organization of the classification system, and I will focus here on
issues of physical interpretation. Because of limited space, this
section only touches upon the details of the Hubble system -- in
particular I do not have space to describe the important extensions to
the Hubble sequence developed by de~Vaucouleurs \cite{deV:1976}.  The
reader can consult several excellent reviews for more comprehensive
descriptions of these issues \cite{Kormendy:1982, Sandage:1994,
vdB:1998}.  Furthermore, I will not be able to consider alternative
classification systems in this lecture -- in my view the most
important of these is the Yerkes system developed by Morgan
\cite{Morgan:1958, Morgan:1959}, based on central concentration of
galaxian light.

\begin{figure}[tbph] 
\epsfig{figure=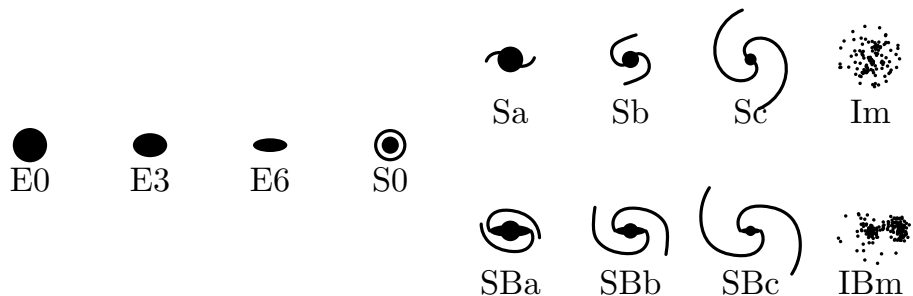,width=12cm} 
\caption{The ``tuning fork'' of the classical Hubble sequence
\cite{Hubble:1926}. As described below, the correlation between
position on this sequence and the physical properties of galaxies
suggests a strong underlying link between morphology and galaxy
evolution.}
\end{figure}

\subsection{Ellipticals}

The sequence E0--E6, based on apparent ellipticity, was originally
conceived of as an angular momentum sequence in young galaxies.
However, it did not take long for colour-based work to suggest that
ellipticals are old \cite{Baade:1963}, and formed quickly in a short
burst of star-formation. Furthermore, it is now known that ellipticals
are pressure-supported by anisotropic velocity distributions, with
rotation playing a subsidiary role (except in less luminous systems
\cite{Davies:1983}).  Since ellipticals are known to be triaxial, the
E0--E6 sequence is seriously contaminated by orientation effects.

The observational evidence supporting the view that ellipticals are
old and roughly coeval is based primarily on the small scatter in the
colour-magnitude diagram of low-redshift {\em cluster} ellipticals
\cite{Sandage:1978, Bower:1992}. This test has also been extended to
higher redshift cluster samples with morphological classifications
from the {\em Hubble Space Telescope} (HST) \cite{Ellis:1997,
Stanford:1998}, where the scatter remains small, suggesting that the
epoch of elliptical formation in rich clusters was well in the past
even at high redshifts.

\subsubsection{Age Controversy}

Although ellipticals have been considered to be old for the last 40
years, the term ``early-type'' to describe these galaxies (based on
the original misapprehensions of Hubble) remains entrenched.  Perhaps
this will turn out to be a blessing in disguise, since the age
distribution of elliptical galaxies, long thought a settled issue, has
recently become enlivened by controversy.  Ellipticals (and S0
galaxies) are predominantly found in dense environments, and
consequently most local work has focussed on the properties of
ellipticals in rich clusters.  But hierarchical models for galaxy
formation introduced the possibility that elliptical galaxies in the
field are continuously formed from merging systems.  In a
controversial paper, Kauffmann {\em et al.}  \cite{Kauffmann:1996}
claim observational evidence, from a colour-based analysis of data
from the {\em Canada-France Redshift Survey} (CFRS), for a factor of
three decrease in the abundance of ellipticals by $z=1$. The claimed
decrease is in good agreement with the predictions of semi-analytical
approximations to hierarchical formation models for galaxy
evolution. Kauffmann {\em et al.} also point out that cluster samples
are ill-suited to testing the hierarchical picture as they represent
accelerated regions so far as structure growth is concerned. It is
quite possible that most of the rich cluster ellipticals are truly old
as their homogeneously distributed colours imply.  The ultimate test
of Kauffmann {\em et al.}'s conjecture lies in the analysis of
ellipticals in field samples. The currently-available samples are too
small for robust results \cite{Schade:1998, Abraham:1998}, although
Zepf \cite{Zepf:1997} claims the absence of red objects at very faint
limits in the HDF is consistent with hierarchical
predictions. Glazebrook {\em et al.}  \cite{Glazebrook:1998} have also
claimed evidence for younger field ellipticals in data from the HST
Medium Deep Survey, on the basis of $I-K$ colours.

\subsubsection{Environmental Effects}

The tendency for ellipticals to be found in clusters is a
manifestation of the {\em morphology-density relationship}, a term
coined by Dressler in the first really thorough investigation of the
effect \cite{Dressler:1980}, although the basic tendency for
ellipticals to be found in rich clusters was known to Hubble, and the
physical significance of this environmental effect was emphasized by
Spitzer \& Baade (1951).  The issue has resurfaced as point of
controversy in recent investigations which seek to determine whether
the key correlations are between morphology and local galaxy density,
or between morphology and radial distance from the centre of rich
clusters \cite{Sanroma:1990, Whitmore:1991}.  Obviously local galaxy
density and cluster-centric radius are strongly correlated, so the
distinction can be rather subtle, but resolving this issue is central
to understanding whether the distribution of galaxy types is built up
gradually from large scale structure by accretion from the field, or
is instead primordial in origin (the {\em nature versus nurture}
controversy).

\subsubsection{Luminosity and the Fundamental Plane}

A remarkable, but perhaps under-appreciated, feature of elliptical
systems is the huge range of luminosity spanned by the class.  The
luminosity of ellipticals spans a dynamic range of around 50,000 (12
magnitudes), from inactive dwarf systems ($M_V \sim -11$ mag) to giant
cD galaxies ($M_V \sim -23$ mag). In contrast, an individual
classification ``bin'' along the spiral arm of the tuning fork
encompasses a range of perhaps 3--5 mag (type Sb having the smallest
dynamic range, and Sd the largest \cite{vdB:1998}).  The enormous
luminosity range encompassed by ellipticals masks an underlying
physical order that is not obvious from the classification
scheme. Ellipticals show a large scatter in quantitative photometric
properties, but as Dressler {\em et al.}  \cite{Dressler:1987} and
Djorgovski \& Davis \cite{Djorgovski:1987} first showed, correlations
between radius, surface brightness, and velocity dispersion in
ellipticals trace out a remarkably well-delineated plane in
three-dimensional parameter space.  Understanding the origin of this
{\em fundamental plane} is a key component in studies of the physical
morphology of early-type systems.  While a large portion of the
scatter in the fundamental plane is due to metallicity variations, it
is important for the student to bear in mind the remarkable fact that
luminous ellipticals are among the most highly enriched galaxies in
the Universe (up to several hundred percent solar metallicity in their
inner portions, with fairly strong internal gradients).  The processes
through which old systems may have undergone rapid enrichment at the
time of formation are fascinating, and well-described by Arimoto
\cite{Arimoto:1997} and references therein.

\subsection{S0 Galaxies}

It interesting to note that on the basis of the early versions of the
tuning fork, S0 galaxies were essentially predicted to exist by Hubble
even before they were observationally established (``The transitional
stage, S0, is more or less hypothetical'' -- Hubble 1936).  The
natural way in which subsequent observations entrenched S0 systems as
key components in the galaxy population has been rightly hailed as a
major success of the Hubble system.  But are S0 galaxies, as predicted
by Hubble, a truly transitional physical ``bridge'' between spirals
and ellipticals?  The evidence is equivocal, and the question remains
among the most interesting in physical morphology.

\begin{figure}
\begin{center}
\epsfig{figure=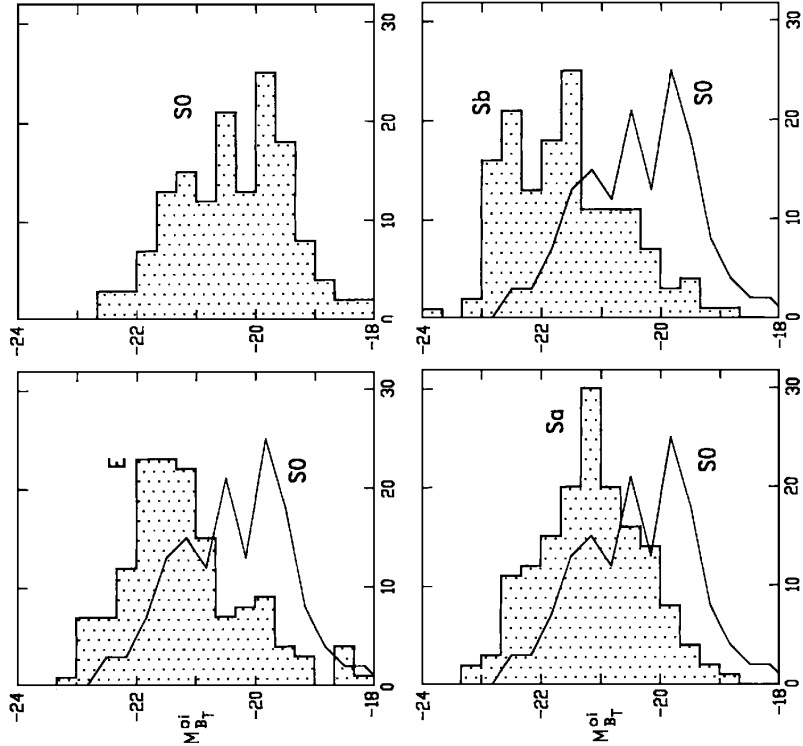,width=10cm,angle=90} 
\caption{Histograms showing the luminosity distributions of
elliptical, S0, Sa, and Sb galaxies in the {\em Revised Shapley-Ames
Catalog} (Sandage 1981). The S0 distribution is repeated from panel to
panel. Note how S0 systems appears to be systematically fainter than
{\em both} elliptical and Sa galaxies, suggesting that S0 galaxies are
not a morphological bridge between these.  Figure taken from van den
Bergh (1998).}
\end{center}
\end{figure}

As pointed out by van den Bergh (1998), {\em the luminosity
distribution of both ellipticals and early-type spirals peaks $\sim
1.5$ mag brighter than that of S0 galaxies}, suggesting S0 galaxies
are {\em not} a smooth transition class between spirals and
ellipticals (Figure~2).  Other evidence seems to be building that
suggests that S0 systems are relics of spirals.  For example, the
``Morphs'' collaboration \cite{Dressler:1994} have used HST imaging to
show that much the gradual bluing in the cluster population with
redshift (the {\em Butcher-Oemler effect}) is due to fairly-normal
looking late-type spirals that are seen in the cores of high redshift
systems, but which are absent in the cores of local clusters. The
elliptical fraction as a function of redshift seems constant, and the
clear implication is that spirals are being transformed into S0
systems.  Coming at this issue from the other direction, other studies
\cite{Bender:1988, Nieto:1988} have shown that some ellipticals
contain boxy inner disks, and other studies have on the whole lent
support to the notion that early-type galaxies can be physically
differentiated on the basis of having disk components of varying
strengths \cite{Kormendy:1996, Faber:1996, Saglia:1997}, possibly as a
function of luminosity.  An interesting synthesis of these ideas has
been proposed by Kormendy \& Bender (1996), who suggest that the
present ranking of the early-types along the tuning fork (based on
apparent eccentricity) should be replaced by a sequence in which
ellipticals are ranked according to boxiness of isophotes
\cite{Kormendy:1996}.  S0 systems would then indeed form a
transitional bridge between spirals and ellipticals with boxy
isophotes. This proposed system is a radical alteration of the Hubble
sequence, but one which preserves the spirit of the S0 class as a
transitional bridge as envisioned by Hubble.

\subsection{Late-type Galaxies}

In Figure~1 spirals are subdivided into barred and ordinary spirals
--- not ``normal spirals'', as they are sometimes termed.  In fact around
$\sim$ 65\% of spirals have recognizable bars or bar-like features in
the $B$-band, according to de~Vaucouleurs (1963).  Along the tines of
the tuning fork the spirals are further subdivided into classes
ranging from early to late-type (Sa, Sb, etc) based on three criteria:
(a) the dominance of the bulge; (b) the degree of winding of the arms;
and (c) degree of resolution of the arms
\footnote{It is rather unsatisfactory to have the early-late spiral
sequence based on three criteria, some of which may be contradictory,
rather than on quantitative measures.  For example, how does one
classify a spiral with a smallish bulge and tight spiral structure (a
combination that is not unknown). Is this designated as an early type
(based on arm appearance) or late-type (based on bulge prominence), or
perhaps classed as peculiar? The answer is that in these case one must
look carefully through the {\em Hubble Atlas} for similar objects and
determine what Sandage chooses to call such a galaxy -- the system is
subjective and ultimately defined by reference to archetypes.}.  A
major weakness of the Hubble system is that the variety of observed
spiral structure is poorly encompassed by the system
\cite{Kormendy:1982}.  Important classes are missing, such as the
flocculent spirals \cite{Elmegreen:1982}, and aenemic spirals
\cite{vdB:1976} (these latter systems are seen only in clusters, and
exhibit faint, ``ghostlike'' spiral structure, suggesting they may be
related to S0 galaxies). In fact this deficiency led van den Bergh
\cite{vdB:1960} to propose the important luminosity class extension to
the Hubble system, in which arm morphology/surface brightness is used
to rank spirals.  This system roughly tracks galaxy luminosity,
exploiting the fact that most luminous spirals have well-developed,
long spiral arms ({\em grand design} spiral structure), while lower
luminosity systems tend to exhibit poorly developed, disconnected
arms.

\subsubsection{Two physical regimes along the tuning fork?}

In probing the physical meaning of the Hubble sequence it is perhaps
useful to mentally subdivide the sequence into two regimes: E--Sbc and
Sc--Irr \cite{vdB:1998} since in the latter objects the dominant
physics seems to be an almost perfect sequence of decreasing mass and
luminosity with Hubble type.  For very late-type galaxies,
correlations between morphology and colour/star-formation history
exist but are not strong -- a typical Im is not significantly bluer
than a typical Scd \cite{Roberts:1994}.  It is perhaps worth noting
that while the general trends in very late-type systems are clear,
there appears to be no really objective way of distinguishing between
very late-type spirals (ie.  types Sd and Sm) and irregular galaxies
\cite{vdB:1998}.

Along the earlier portion of the sequence the Hubble system remains
quite successful at ranking galaxies along physical lines, although
the story is more complex.  In addition to ranking galaxies by
bulge-to-disk ratio (partly by definition), the early part of the
spiral tines of the Hubble sequence order galaxies by surface HI and
total mass density, total HI mass, colour, and, to some extent,
mass-to-light ratio.  Of these the strongest correlations appear to be
with colour \cite{Roberts:1994}, and hence with star-formation
history.  To what extent these correlations would remain if one ranked
galaxies simply by bulge-to-disk ratio, as proposed by Morgan
\cite{Morgan:1958,Morgan:1959}, is unknown.  However, because
bulge-to-disk ratio is the dominant morphological characteristic
linked to star-formation history, and because this seems the strongest
correlation, I would not be surprised if the correlations along the
early part of the spiral sequence actually improved if the ranking
was purely by bulge-to-disk ratio.

\subsubsection{Bulge formation}

The physics of bulge formation in spirals is of great interest.  In
our own Galaxy the observational evidence for early bulge formation
seems established, mainly since the main tracers of bulge/halo
populations (eg.  globular clusters and RR Lyrae stars) are known to
be old.  The notion that bulges form well before the discs in the
first stage of galactic collapse is the essential component in both
the classic Eggen, Lynden-Bell, \& Sandage (ELS) scenario
\cite{Eggen:1962} and more modern variants \cite{Carney:1990}.
Hierarchical galaxy formation scenarios also require old bulges formed
by mergers \cite{Kauffmann:1993, Baugh:1996}, with visible discs
built-up gradually from gas accreted onto these merger
remnants. Recently, however, the issue of the relative age of the
bulge and disc in extragalactic systems has become controversial.
N-body simulations indicate that bulges form naturally from bar
instabilities in discs \cite{Norman:1996, Combes:1990}, and recent
observations now seem to indicate that bulges display the
morphological and dynamical characteristics of such a formation
scenario (eg. triaxiality or ``peanut'' shapes \cite{Kuijken:1995},
cylindrical rotation \cite{Shaw:1993}, and disky kinematics
\cite{Kormendy:1992}).  Most remarkably, recent observations indicate
that the inner discs and bulges in local galaxies {\em cannot be
distinguished in terms of colour} \cite{Balcells:1994,deJong:1996},
which is surprising given the visual impression from ``true colour''
deep HST images of distant galaxies, in which bulges seem generally
redder than disks \cite{vdB:1996}. Recent quantitative modelling of
low-redshift spirals in the Hubble Deep Field \cite{Abraham:1998} also
seems to suggest that bulges are generally the oldest parts of
galaxies\footnote{However, if secular activity somehow generates
bulges from disks {\em without} any star-formation activity (ie. by
simply reorganizing existing galaxy populations without forming any
new ones), then it is possible that red bulges could still be formed,
provided they were built-up exclusively from the oldest disk
populations in the centres of the galaxies. The plausibility of this
hypothesis could perhaps be tested by hydrodynamical simulations. I am
grateful to Mike Merrifield for pointing out this possibility.}.

\subsection{An improved Hubble system?}

Most observers would agree that (locally) the Hubble system supplies a
fairly complete framework for classifying galaxies in the field -- it
seems that over 90\% of local field galaxies find a natural home
within the system.  Furthermore, discussion of the previous section
argues rather forcefully for a close connection between the
morphological bins of the Hubble system and an underlying physical
order.  The major criticism I would level against the ranking implicit
in the Hubble sequence is that the classification of the early-type
portion of the diagram (based on apparent eccentricity) seems to me
rather unphysical.  Furthermore the system is biased heavily in favour
of luminous giant and supergiant galaxies (which dominate the galaxian
mass content of the Universe, but which do not dominate in terms of
total numbers seen in volume-limited samples). My own view is that
both of these criticisms would be greatly alleviated if luminosity
were included as a classification criterion. This is not really
practical at the present time, since redshifts are rarely available
for random galaxies on CCD images.  But at least as an organizational
structure, it seems to me that a three-dimensional system such as that
proposed by van den Bergh (1998) has much to commend it (Figure 3),
summarizing pictorially many of the issues dealt with so far in this
lecture.

\begin{figure}[tbph] 
\begin{center}
\epsfig{figure=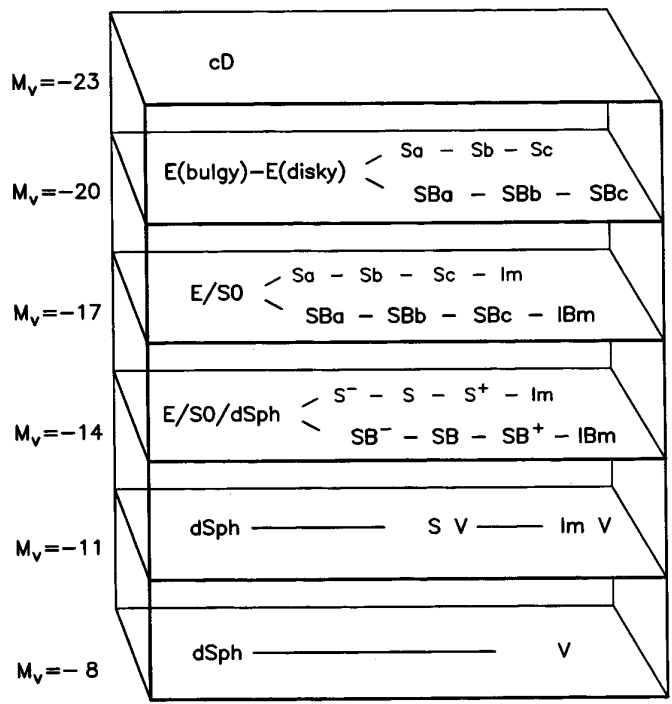,width=9cm} 
\caption{The three-dimensional tuning fork proposed tentatively by van
den Bergh \cite{vdB:1998}. Note how the the early-type sequence has
been replaced in all panels by more physically meaningful
classifications. As one descends the luminosity scale of the
three-dimensional tuning fork, the conventional forms of late-type
galaxies on the Hubble sequence disappear, and become replaced by
lower-luminosity systems with more ragged morphology.}
\end{center}
\end{figure}

On balance, I would argue that according in terms of the first two of
the three (rather strict) criteria for a useful classification scheme
espoused in \S2.1, the Hubble system is a resounding success.
However, because of the rather depressing results from controlled
comparisons between independent classifications made by expert
morphologists \cite{Naim:1995}, I have rather grave doubts about the
third criterion (observer-to-observer consistency in visual
classifications --- I have no doubt that an expert morphologist can
make classifications that are, internally, highly consistent). Recent
years have seen great advances in the usefulness of objective,
machine-based morphological classifications \cite{Doi:1993, Naim:1995,
Abraham:1996a, Odewahn:1996}.  None of these automated classification
schemes is able to reproduce the Hubble system in detail, although the
classifications do track the system in a crude sense. At high
redshifts, where low signal-to-noise and subtle selection effects play
havoc with visual classifications, simple and objective
classifications that can be calibrated by simulations are essential.

\section{MORPHOLOGY AT HIGH REDSHIFTS}

As one probes further into redshift space, it would be surprising if
the Hubble sequence did not begin to break down as one approaches the
initial epoch of galaxy formation.  The interesting questions are {\em
where} the system breaks down, and {\em how}. Does one class of galaxy
within the system begin to gradually dominate over the overs,
indicating that the sequence itself contains the ``galaxian ground
state'' as one of its classification bins?  Or do entirely new classes
of galaxy emerge?  With the advent of HST, we are now in a position to
address such questions.

Recent work from deep HST imaging surveys
\cite{Griffiths:1994,Glazebrook:1995,Driver:1995,Abraham:1996a,
Abraham:1996b,Giavalisco:1996} coupled with ground-based spectroscopic
work, \cite{Lilly:1995,Cowie:1995,Lilly:1996,Ellis:1996} has shown
that much of the rapidly evolving faint blue galaxy population
\cite{Broadhurst:1988,Koo:1992,Lilly:1995} is comprised of
``morphologically peculiar'' galaxies. This term has been rather
liberally applied to encompass a vast range in observed galaxy forms,
but in fairness more precise classifications have been difficult to
apply, because at high redshifts galaxies are being observed in the
rest-frame ultraviolet (a ``morphological K-correction''), where
little is known about the appearance of the local galaxy
population. However the conclusion that these systems are {\em
intrinsically} peculiar seems secure, because the general effects of
cosmological bandshifting on normal Hubble types has been determined
from simulations \cite{Abraham:1996b}. In general the observed faint
peculiar systems do not resemble the appearance of bandshifted Hubble
sequence galaxies.  Furthermore the redshift range probed by most deep
$I$-band HST imaging corresponds to $z<1$ (with the exception of
Lyman-limit selected systems discussed below), in which the effects of
cosmological bandshifting on morphology are not yet extreme.

\begin{figure} 
\begin{center}
\epsfig{figure=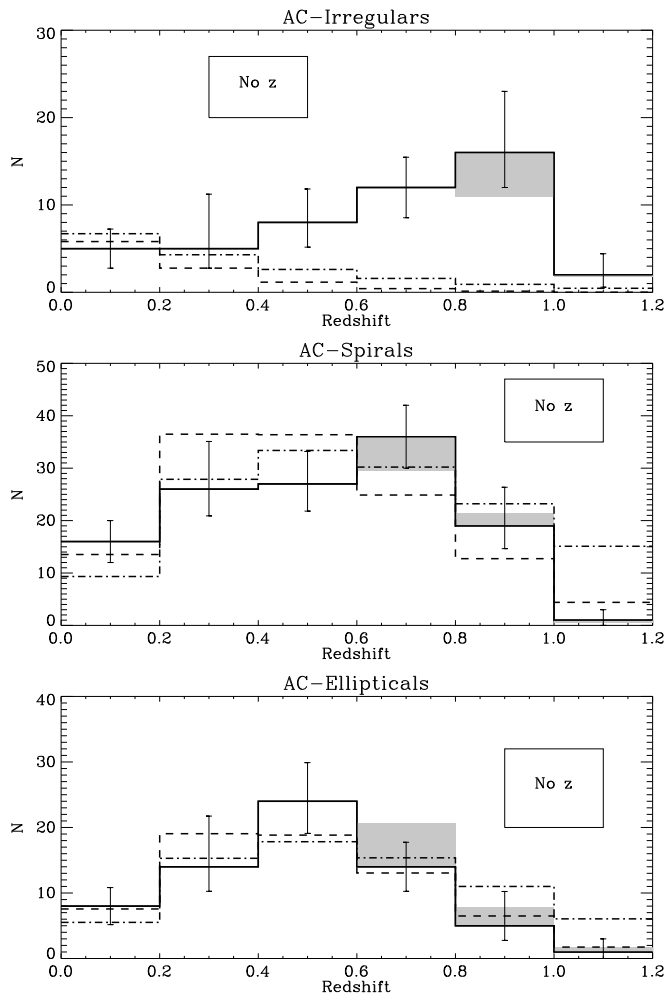,width=9.5cm}
\caption[NZ]{Morphologically segregated number counts from Brinchmann
{\em et al.} (1998) \cite{Brinchmann:1998}, based on data from the
CFRS/LDSS collaboration. The solid-line bins show counts as a function
of redshift for irregular/peculiar/merger systems (top), spirals
(middle), and ellipticals (bottom). Morphological classifications have
been made from WF/PC2 images using an automated technique based on
central concentration and asymmetry of galaxian light
\cite{Abraham:1996b}.  The shaded region corresponds to the size of
the ``morphological K-correction'' on the classification, accounting
for the effects of observing the galaxies at bluer rest wavelengths as
a function of redshift. Superposed on the observed histograms are the
predictions of no-evolution (dashed) and 1 mag linear evolution to
$z=1$ (dot-dashed) models. At $z\sim 1$ approximately 40\% of the
galaxy population is morphologically peculiar.}
\label{fig:nz}
\end{center}
\end{figure}

Perhaps the clearest evidence for the increasing importance of
morphologically peculiar systems as a function of redshift has been
obtained by Brinchmann {\em et al.} \cite{Brinchmann:1998}.  These
authors applied an objective classification scheme, calibrated by
simulations, to a set of $\sim 300$ HST $I_{814}$-band images of
galaxies {\em with known redshifts} taken from the CFRS
\cite{Lilly:1996} and LDSS \cite{Ellis:1996} surveys. Because the
statistical completeness of this sample is very well understood,
reliable number-redshift histograms can be constructed for the various
morphological types. The morphologically resolved $n(z)$ result
obtained by Brinchmann {\em et al.} is shown in Figure~4, and confirms
that irregular/peculiar/merging systems are already greatly in excess
of the predictions of no-evolution and mild-evolution models at
redshifts $z\sim1$.  It is clear that by $z\sim1$ approximately 1/3 of
galaxies are morphologically peculiar.

What are these peculiar systems? The answer to this question is
currently unknown. These galaxies are often referred to as
``irregulars'' in the literature, but it is probably a mistake to
regard these systems as counterparts to local irregulars.  As pointed
out in the previous section, luminous irregulars are virtually unknown
in the local Universe, while the high-redshift peculiar systems are
generally both large and bright\footnote{It is left as an exercise for
the reader to show that the selection function for a magnitude-limited
survey sampling a Schechter luminosity function results in a roughly
gaussian-shaped absolute magnitude distribution that peaks near
$L_\star$}.  

Let me conclude this section by pointing out that although (because of
space limitations) my focus in these lectures is on peculiar galaxies
at high redshifts, one should not lose sight of the importance of
tracking systematic changes in the characteristics of morphologically
{\em normal} systems. An interesting study has recently been completed
by Lilly and collaborators \cite{Lilly:1998}, which seems to indicate
little change in the space density of large spiral systems to redshift
$z=1$. The distribution of galaxian disk sizes is a sensitive probe of
hierarchical formation scenarios (in which disk sizes are expected to
strongly evolve with redshift). Attempts to understand the
implications of this observation in the context of hierarchical models
are underway \cite{Mao:1998}.

\subsection{The nature of high-z peculiar galaxies} 

\subsubsection{The evolving merger rate}

Locally, most morphologically peculiar systems show dynamical evidence
for tidal disruption, and it is tempting to assume that a large
fraction of the diverse peculiar galaxy population seen on deep images
are actually mergers in progress. But is this assumption justified?
Figure 5 shows candidate $z>2$ Lyman dropout systems in the Hubble
Deep Field with $I_{F814W} < 25$ mag. Clearly most are morphologically
peculiar, but few resemble the classical appearance (ie. strongly
bimodality, with prominent tidal tails) of ``canonical'' local merging
systems.  However, as discussed further below, the $z>2$ regime
accessible to Lyman limit searches probes rest wavelengths where the
effects of bandshifting on morphology can be rather extreme.  This is
made clear by Hibbard \& Vacca \cite{Hibbard:1997}, who used HST FOC
ultraviolet data of local merger-induced starburst galaxies to predict
the appearance of the high-redshift counterparts. The usual signatures
of mergers (tidal tails, distorted disks) are no longer visible at
$z>2$, and merging starbursts seem to provide at least qualitatively
reasonable counterparts to many faint peculiar galaxies.

\begin{figure} 
\begin{center} 
\epsfig{figure=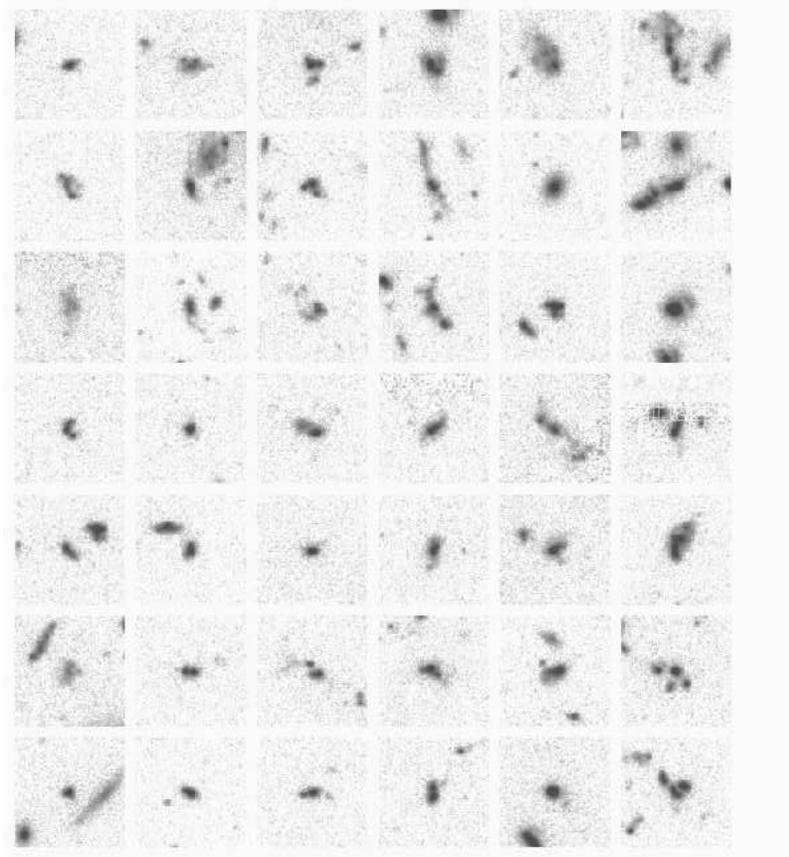,width=4.5in,angle=-90} 
\end{center} 
\caption{Candidate lyman break ($z>2$) systems in the Hubble Deep
Field with $I<25$ mag, taken from van den Bergh {\em et al.}
(1996) \cite{vdB:1996}.}
\end{figure}

Clearly the best way forward will be to incorporate dynamical
information to determine directly which peculiar galaxies show
distinct kinematical subcomponents. Unfortunately these observations
are not currently feasible, although they may soon become possible
with adaptive optics and the new generation of 8m-class telescopes. In
the meantime, a promising approach to quantifying the fraction of
mergers amongst the distant peculiar galaxy population may be to
measure statistics which are relatively insensitive to image
distortions resulting from bandshifting and surface-brightness biases,
but which track probable merger activity. One such statistic is the
``Lee Ratio'', a measure of image bimodality. This statistic been
applied to images of galaxies in the CFRS survey
\cite{LeFevre:1997,LeFevre:1998} and to HDF galaxies, with the result
that around $\sim 40\%$ of faint peculiar systems are significantly
bimodal, with an $\sim (1+z)^3$ increase in the merger rate.

\subsubsection{When does a merger become a galaxy?}

The observed evolution in the merger rate, coupled with the
expectation from theory that of mergers move galaxies up and down the
Hubble sequence (cf. the lectures given by Prof. White in this
volume), forces morphologists studying the high-redshift Universe to
confront both philosophical and practical issues that seemed rather
semantic and trivial at low   redshifts.  For example: when should an
amorphous blob of components be regarded as single morphologically
peculiar galaxy, as opposed to a system of interacting proto-galaxies?
de~Vaucouleurs used to dismiss the notion of considering mergers to be
fundamental morphological units with the observation that ``car wrecks
are not cars''. But when the road is littered with wrecks, and when
the by-product of a wreck is another working car, it may be time to
re-assess the wisdom of restricting morphological classification to
regular-looking systems.  Classification schemes which explicitly
incorporate measures of asymmetry and bimodality are needed to
realistically capture the appearance of galaxies in the high-redshift
Universe.

\begin{figure} 
\begin{center} 
\epsfig{figure=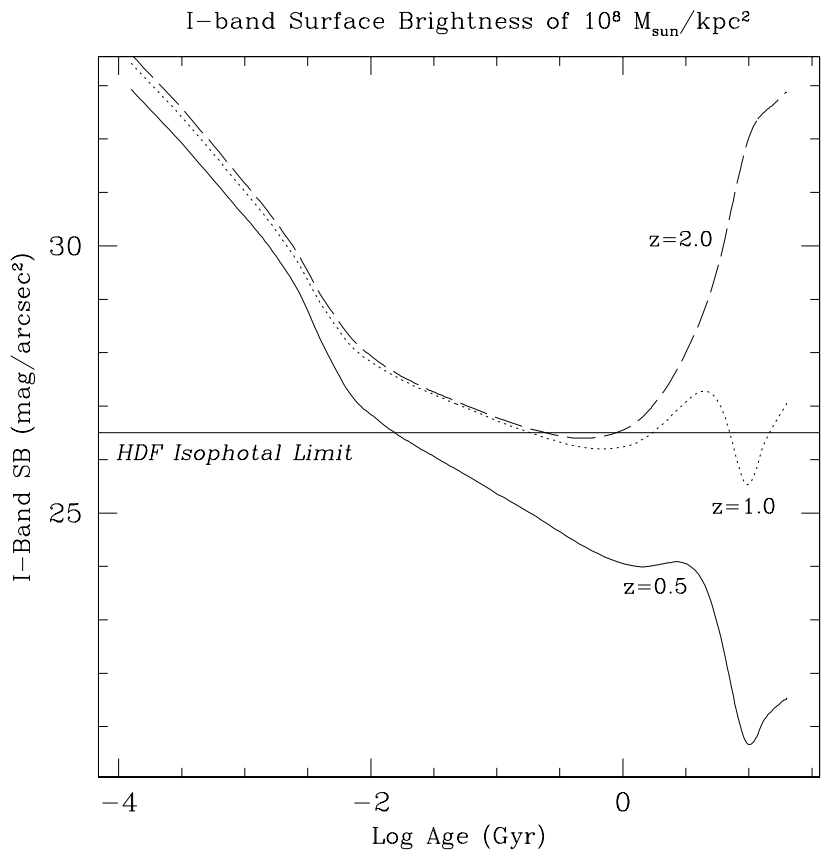,width=4.5in} 
\end{center} 
\caption{Predicted $I$-band surface brightness as a function of age
for a stellar population with a projected surface mass density of
$10^8$ $M_\odot$ kpc$^{-2}$. The three curves shown correspond to
stellar populations with exponential star formation histories (with
e-folding timescales of 1 Gyr) seen at redshifts $z=0.5$, $z=1.0$, and
$z=2.0$. Calculations are based on the predictions of the GISSEL96
spectral synthesis package (Bruzual \& Charlot 1996).  Also shown is
the approximate $I$-band (F814W) isophotal detection limit for the
Hubble Deep Field. Note the precipitous decline in visibility of older
stellar populations with redshift. Curves such as these suggest that
beyond $z>1.5$ substantial mass could be locked in ``old'' (ages
greater than one or two Gyr) stellar populations that would be
undetectable even in red optical bands.}
\end{figure}

\subsubsection{Morphologies of Lyman-limit systems considered
dangerous}

Deep HST imaging of galaxies selected on the basis of colour to be at
redshifts $z>2$ has provided an important breakthrough in our
understanding of high-redshift morphology \cite{Steidel:1996,
Giavalisco:1996}.  Such surveys probe systems seen at ages
corresponding to 10-20\% of the present age of the Universe. So far
the results appear somewhat contradictory. Early studies focussed on
the compact cores of these systems, which appear well-modelled by
$r^{1/4}$ law profiles, suggestive of proto-ellipticals (Giavalisco
{\em et al.} 1996). Deeper observations have revealed morphological
characteristics among the most bizarre yet seen on deep HST data
(Fig~5). Whether the morphologies of Lyman limit systems are
interpreted as fairly regular or totally bizarre is a strong function
of the limiting surface brightness of the observations, and whether
the complex structures linking the bright knots have been
resolved. Thus any interpretation of the morphologies of Lyman limit
galaxies remains speculative. In the context of these lectures perhaps
the best that can be done is to present explicitly the various factors
that make the interpretation of the $z>1.5$ morphological data so
subtle.

The most obvious complication is that Lyman break systems are being
observed at rest wavelengths so far in the ultraviolet that even the
relatively sparse ground-based $U$-band data of local systems provide
a poor reference standard for comparison. (In fact, the handful of
local systems observed with the {\em Ultraviolet Imaging Telescope} on
the Astro-1 and Astro-2 missions provide the best calibration
reference.)  In any case, Lyman limit systems are at redshifts so high
that the effects of evolution {\em must} be incorporated explicitly
when making meaningful comparisons. At the epoch being observed, the
Universe may simply be too young to have evolved the old stellar
populations that play an important role in defining local
morphology. Also, extraordinarily strong surface-brightness selection
effects bias against the detection of even intermediate-aged stellar
populations. These latter points are perhaps best understood from
plots such as that shown in Figure~6, which illustrates the surface
brightness of a surface mass density as a function of population age
and observed redshift.  Stellar populations with ages greater than
1--2 Gyr are strongly biased against in observations at $z>1.5$, and
only the youngest stellar populations (the ``tip of the iceberg'' in
terms of stellar mass) are detectable. On the other hand, if $\Omega$
is large then the Universe at $z \sim$ 2--3 may be sufficiently young
that the populations biased against would not have time to have
formed, so a large proportion of the total mass would be detectable.

\subsubsection{New classes of galaxies}

The possibility that many morphologically peculiar high-redshift
galaxies may be more-or-less conventional mergers ({\em ie.}
interactions between established galaxies) should not blind us to the
likelihood that many of the observed galaxy forms probably correspond
to entirely new classes of objects, or perhaps to the initial merging
events forming the first generation of luminous galaxies.  For
example, it has been pointed out that a fairly large proportion of
faint peculiar systems have knotty, linear forms
\cite{Cowie:1995}. Several examples are seen in Figure 5. The nature
of these ``chain galaxies'' is controversial. Some authors suggest
that they may be edge-on spiral or low surface-brightness disk systems
\cite{Dalcanton:1996}.  However, I believe the evidence (at least for
the most striking chain galaxies) strongly suggests that these deserve
to be considered a {\em bona fide} new class of object.  For example,
the colours of the knots in some relatively low-redshift chains are
remarkably synchronized --- giant complexes of star-formation appear to
propagate along the body of these system like a string of fireworks
\cite{Abraham:1998b}. While intrinsically linear systems are
dynamically unstable on timescales $\sim 100$ Myr, a straight-forward
comparison of the internal colours of these galaxies with spectral
synthesis models suggests that the unweighted mean age of the
starlight in chain galaxies is indeed of order 100 Myr, and thus
comparable with the dynamical timescale. The age difference between
the youngest and oldest knots in the chain is only around 30--50 Myr.
If these properties prove universally true (and one needs to be
cautious, as only a few chain galaxies have been examined in detail so
far), then chain galaxies certainly seem to be systems with no local
analogue.

How many other entirely new classes of peculiar galaxy exist at high
redshifts? We will not know until such systems can be distinguished
from more straight-forward mergers (perhaps through their dynamical
properties), and until an inventory of the different classes of
peculiar systems is undertaken over a broad range of redshifts.  Until
such studies are undertaken it will be difficult to disentangle the
physical mechanisms responsible for the myriad spectacular forms of
the distant galaxies seen on deep HST images.

\vspace{1cm}

\noindent{\bf Acknowledgments} I am grateful to Olivier Le F\`evre and
the other organizers of this school for giving me a welcome excuse to
spend many pleasurable hours studying the {\em Hubble Atlas}. I thank
my collaborators Richard Ellis, Jarle Brinchmann, Karl Glazebrook,
Andy Fabian, Sidney van den Bergh, Nial Tanvir, and Basilio Santiago
for their many contributions to the projects described in this
article. I am also grateful to Simon Lilly and the rest of the CFRS
team for useful discussions, and for permission to describe results in
advance of publication. I also thank Mike Merrifield for interesting
discussions on the ages and colours of bulges.

\end{document}